# Development of a novel hybrid haptic (nHH) device with a remote center of rotation dedicated to laparoscopic surgery.

Majdi Meskini†‡, Houssem Saafi†§, Abdelfattah Mlika†,

Marc Arsicault‡, Said Zeghloul‡, and Med Amine Laribi‡

†Mechanical Laboratory of Sousse (LMS), National Engineering School of Sousse, University of Sousse, Sousse 4000, Tunisia.

‡ University of Poitiers, CNRS, ISEA-ENSMA, Pprime Institute, UPR 3346, Poitiers, France.

§Preparatory Institute for Engineering Studies of Gafsa, University of Gafsa, Gafsa 2000, Tunisia.

**Abstract**

This paper focuses on developing a novel Hybrid-Haptic (nHH) device with a remote center of rotation with 4 DOFs (degrees of freedom) intendant to be used as a haptic device. The new architecture is composed of two chains handling each one a part of the motions. It has the advantages of a parallel robot as high stiffness and accuracy, and the large workspace of the serial robots. The optimal synthesis of the nHH was performed using real-coded Genetic Algorithms (GA). The optimization criteria and constraints were established and successively formulated and solved using a mono-objective function. A validation and comparison study were performed between the spherical parallel manipulator and the nHH. The obtained results are promising since the nHH is compared to other similar task devices, such as (SPM), and presents a suitable kinematic performance with a task workspace free singularity inside.

## 1. Introduction

Cooperation between humans and robots plays an important role in many fields. In fact, haptic devices[1], [2] are developed in order to enable its users to interact with software program or a virtual item by giving a force and torque feedback. They are used to increase the application immersion such as gaming [3], [4], training in virtual environment [5]–[7] and augmented reality [8]. Minimally Invasive Surgery (MIS) is one of practical example where haptic feedback could be very useful[9].

Haptic devices have been widely investigated, Van den bedem proposed in [10] a spherical serial master haptic device with 4-DoFs. The serial architecture has a simple kinematic. However, it has major drawbacks such as all actuators must be placed on the joint axes which increase the required torques and the weight of the end-effector that the surgeon is asked to support. Several other authors [11] proposed a spherical parallel manipulator (SPM) with a remote center of rotation (CoR) used as a haptic device for MIS (Fig1.a). Contrary to the serial master haptic device, actuators of the SPM are located at the base which reduces the inertia issue of the system. However, this haptic device with parallel architecture suffers from singularities inside his workspace, which causes the loss of DoF and the amplification of errors in the kinematic transformation [12]. In order to cope with the singularity problem inside the workspace of a parallel device, Saafi et al. [13] proposed several solutions. Indeed, the behavior of the manipulator is improved by the use of a redundant actuator, the use of an extra sensor as well as a specific control scheme. As reported by Saafi [14] the use of an extra sensor placed in a passive joint improve the calculation of the forward kinematic model even in singular positions. Another solution is described in [15], [16] which introduce a serial approach for solving the forward kinematic model by placing three sensors on one leg rather than placing them on the actuated joint located at the base. This solution aim to improve the calculation of the forward kinematic model and eliminate the parallel singularity effects in real time application.

Another interesting kinematic of a haptic device was presented by Pérault et al in [17]. This architecture is based on Delta robot architecture (Fig 1.b). However, the remote CoM of the mechanism is at the bottom of the robot which increases the gravity compensation and required actuators size.

Hybrid devices as presented in [18]–[21] started to take their places in a lot of fields. A new hybrid haptic device (Fig 1.c) is presented by Saafi et al. [22]. This haptic device is the association of a parallel chain and a serial chain. In this latter, the parallel chain is responsible for the tilt motions and the serial chain handles the self-rotation and the translation. The limitation of this architecture is that the actuator handling the self-rotation should be supported by the serial chain, which increases the weight of the translational part and as consequence the linear actuation torque.

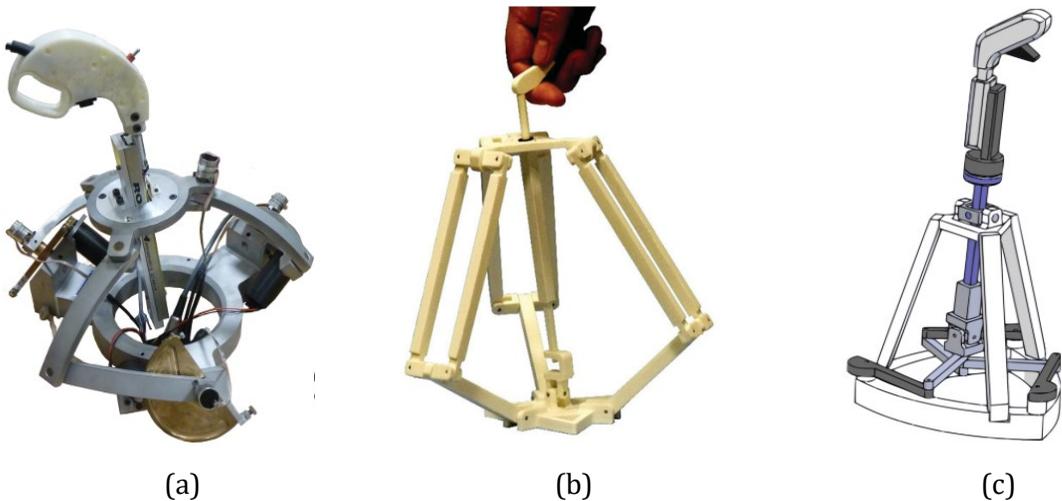

(a) (b) (c)

**Figure 1.** *(a): SPM haptic device*[23] *(b): haptic device based on Delta robot*[17] *(c): hybrid haptic device for laparoscopic surgery*[22].

In this paper, a new haptic device with a remote CoR is designed for MIS based on hybrid architecture for laparoscopic surgery. The proposed architecture is based on the association of a parallel part with 3-RRR (R : revolute) parallel planar manipulator, and a serial part connected through universal joint. The self-rotation is supported by the parallel part. The haptic device is optimized in order to guaranty the best kinematic behavior and torque feedback. Simulations are carried out in purpose to validate the manipulator efficiency and handles comparison with other haptic devices.

The paper is organized as follows. In section 2, the workspace of the desired task will be identified based on the capture of medical expert motions performing MIS. The proposed architecture of the new Hybrid Haptic (nHH) device is discussed on section 3. Section 4. Is devoted to the dimensional synthesis of the manipulator . In section 5. A numerical validation and a comparison between the new HH device and the SPM are caried out. Section 6 summarizes this paper.

## 2. Laparoscopic workspace identification

MIS surgery tools are designed to enter the patient's body through incisions, requiring less time for recovery and less pain for the patient. In order to identify the workspace swept by the instruments during this surgical operation, motion capture system was used to record the gestures of a skilled surgeon [24]. An instance of workspaces deduced from the recorded instruments motions, a clamp and a needle holder, is shown in Fig.2. Each tool works in a space with a conical geometric form defined by a half-vertex angle $\alpha$ that can go up to 26°.

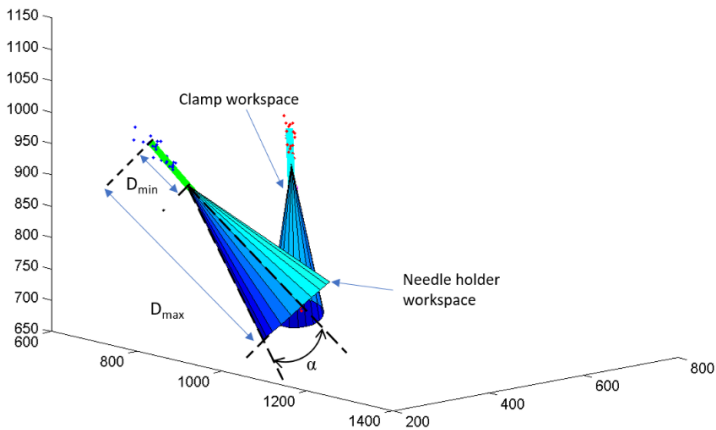

*Figure 2. Experimental MIS workspace.*

Further, each tool moves along the cone's axis with a translational motion defined by its bounding limits given by the maximal and minimal distances (Dmax and Dmin) between the incision point and the tool tip, respectively.

Based on the surgical gestures analysis reported in [25], [26] ,[27], the MIS procedure requires four degrees of freedom, namely three rotations and one translation. One can conclude that the tool workspace is described by a cone with a maximum vertex angle of 26° and a translation of 112mm along the tool axis direction, respectively.

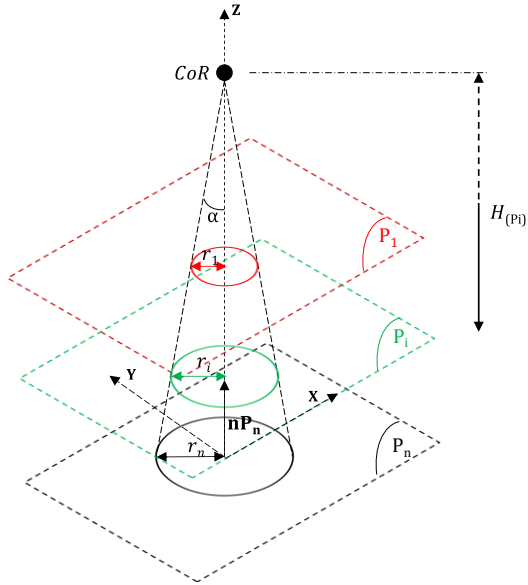

**Figure 3.** *Geometric description of the laparoscopic workspace*

The geometric construction of the tool workspace given in Fig.3 shows the location of the CoR, at the top of the cone, and the intersections with a set of parallel planes $P_i$ (i ∈ {1, 2…, n}) defined by their common normal vector **nP$_n$**. Each plane $P_i$ is located at the distance $H_{p_i}$, between the plane $P_i$ which design the plane of the parallel chain and the CoR. The intersections are described by a set of circles with radius $r_i$ and given by Eq. (1):

$$r_i = H_{(Pi)} \cdot \tan \alpha \qquad (1)$$

The next section will introduce the kinematic of the new haptic device for the MIS, called hybrid haptic device.

## 3. nHH-device

The new proposed architecture, the novel hybrid haptic (nHH) device offers four possible motions as presented in the Fig.4. One can note three rotations around a defined CoR and one translation. This architecture is composed of a serial and a parallel kinematic chain. Each of these chains operate a part of the device. The parallel chain is composed of a 3-RRR planar manipulator with 3 DoFs, which allows to handle the two tilt motions and the self-rotation. The serial chain is composed of a prismatic and spherical joint, allowing to manipulate the translation around a fixed CoR. The two chains of this device are connected through a universal joint. This association allows to overcome with the drawbacks of the serial architectures where the actuators are placed on the joint axes and increase the moving masses which alters dynamic behavior.

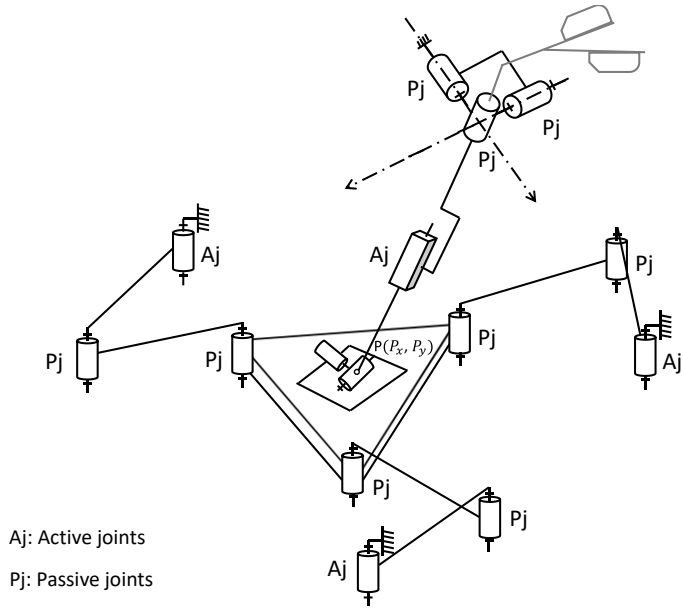

*Figure 4.* nHH device kinematics

In the section below, the two kinematic chains of the presented architecture will be discussed in detail. The CAD model of the nHH device and its developed prototype are presented in Fig.5a. and 5.b, respectively.

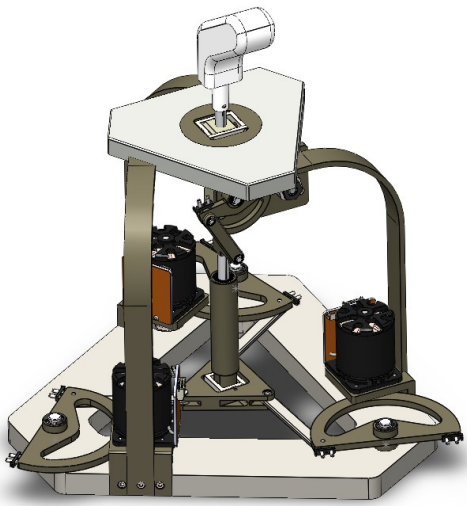
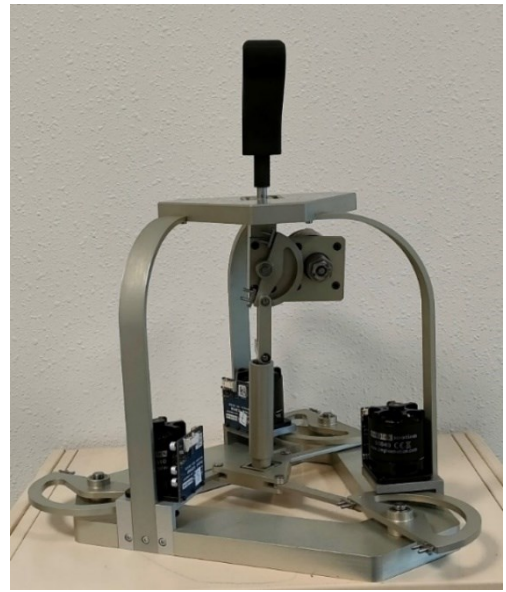

***Figure 5.*** *(a): CAD model of the nHH device*          *(b): nHH device prototype*

## 3.1. Serial chain (SC)

A universal joint serving as a CoR, a revolute joint for self-rotation, and a prismatic joint to control the linear displacement defined by T make up the serial chain of the proposed nHH device. As shown in Fig. 6.(a), the orientations of the serial chain are specified using cardon angles with three different rotations in three-dimensional space. The bond graph of the serial chain is presented in Fig.6.(b).

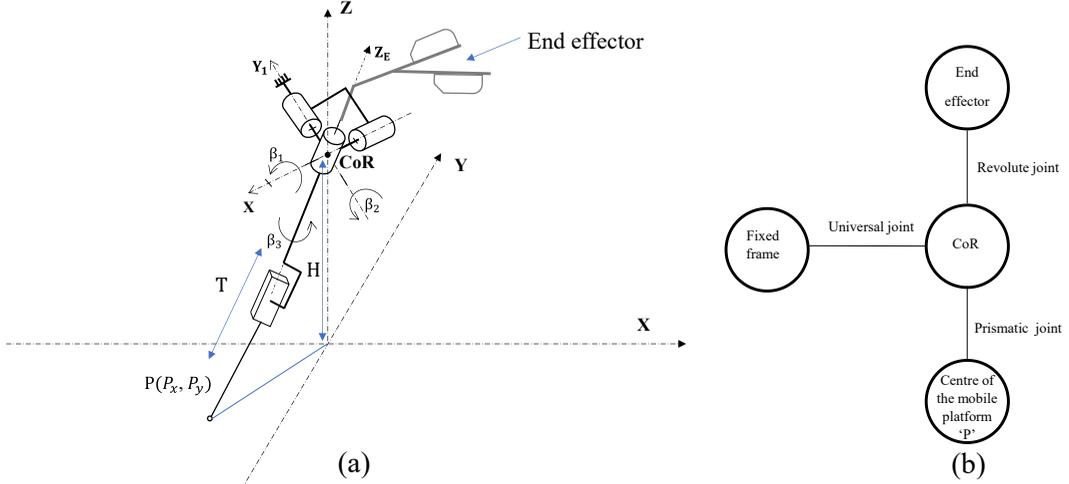

**Figure6.** *(a):Serial chain architecture of the nHH device (b) serial chain bond graph*

The end effector velocity can be expressed using the universal joints $(\beta_1, \beta_2, \beta_3)$ presented in the Fig.6 by Eq.2:

$$\boldsymbol{\omega}_{EF} = \dot{\beta}_1 \mathbf{X} + \dot{\beta}_2 \mathbf{Y_1} + \dot{\beta}_3 \mathbf{Z_E} \tag{2}$$

$$\begin{pmatrix} \omega_x \\ \omega_y \\ \omega_z \end{pmatrix} = \mathbf{Js} \begin{pmatrix} \dot{\beta}_1 \\ \dot{\beta}_2 \\ \dot{\beta}_3 \end{pmatrix} \tag{3}$$

The orientation of end effector in workspace can be also described by the Euler angles (ψ,θ,φ) with ZXZ convention. In this case the angular velocity of the end effector can be expressed as follows:

$$\boldsymbol{\omega}_{EF} = \dot{\psi}\mathbf{Z} + \dot{\theta}\,\mathbf{X_1} + \dot{\varphi}\,\mathbf{Z_E} \tag{4}$$

$$\begin{pmatrix} \omega_x \\ \omega_y \\ \omega_z \end{pmatrix} = \begin{bmatrix} 0 & \cos\psi & \sin\theta \sin\psi \\ 0 & \sin\psi & -\sin\theta \cos\psi \\ 1 & 0 & \cos\theta \end{bmatrix} \begin{pmatrix} \dot\psi \\ \dot\theta \\ \dot\varphi \end{pmatrix} \qquad (5)$$

Or From Eq.3 and Eq.5 we can obtain :

$$\begin{bmatrix} 0 & \cos\theta & -\cos\theta \sin\psi \\ 0 & \sin\theta & \sin\theta \\ 1 & 0 & \cos\theta \end{bmatrix} \begin{pmatrix} \dot\psi \\ \dot\theta \\ \dot\varphi \end{pmatrix} = \mathbf{Js} \begin{pmatrix} \dot\beta_1 \\ \dot\beta_2 \\ \dot\beta_3 \end{pmatrix} \qquad (6)$$

$$\begin{pmatrix} \dot\psi \\ \dot\theta \\ \dot\varphi \end{pmatrix} = \begin{bmatrix} 0 & \cos\theta & -\cos\theta \sin\psi \\ 0 & \sin\theta & \sin\theta \\ 1 & 0 & \cos\theta \end{bmatrix}^{-1} \mathbf{Js} \begin{pmatrix} \dot\beta_1 \\ \dot\beta_2 \\ \dot\beta_3 \end{pmatrix} \qquad (7)$$

where,

$$\mathbf{Js} = [\mathbf{X}\ \mathbf{Y}_1\ \mathbf{Z}_E]. \qquad (8)$$

with,

$$\begin{cases} \mathbf{Y}_1 = R_x(\beta_1)\mathbf{Y} \\ \mathbf{Z}_E = R_z(\psi)\ R_x(\theta)\ R_z(\varphi)\ \mathbf{Z} \end{cases} \qquad \begin{matrix}(9)\\(10)\end{matrix}$$

Using the identification between the Euler matrix and the cardan matrix, $\beta_1$ can be expressed using Euler angles by Eq.11 :

$$\beta_1 = \tan^{-1}(\cos\psi \tan\theta) \qquad (11)$$

Using this result, the $\mathbf{Js}$ matrix is expressed by:

$$\mathbf{Js} = \begin{bmatrix} 1 & 0 & \sin(\psi)\sin(\theta) \\ 0 & \cos(\tan^{-1}(\cos\psi \tan\theta)) & -\cos\psi \sin(\theta) \\ 0 & \sin(\tan^{-1}(\cos\psi \tan\theta)) & \cos(\theta) \end{bmatrix} \qquad (12)$$

The matrix $\mathbf{Js}$ of the serial chain, is used in the evaluation of the kinematic performance of the serial chain. This evaluation is based on the dexterity criterion, detailed in section 3.2.2. The dexterity noted $\mu_s$ is equal to the inverse of the conditioning number of the matrix $\mathbf{Js}$ (Eq.13).

$$\mu_s = \frac{1}{K(\mathbf{Js})} \qquad (13)$$

The dexterity distribution in the plane ($\psi$, $\theta$) of the serial chain is presented in Fig.7. One can conclude suitable kinematic performances of the serial chain since the dexterity is higher than 0.5 over the whole workspace.

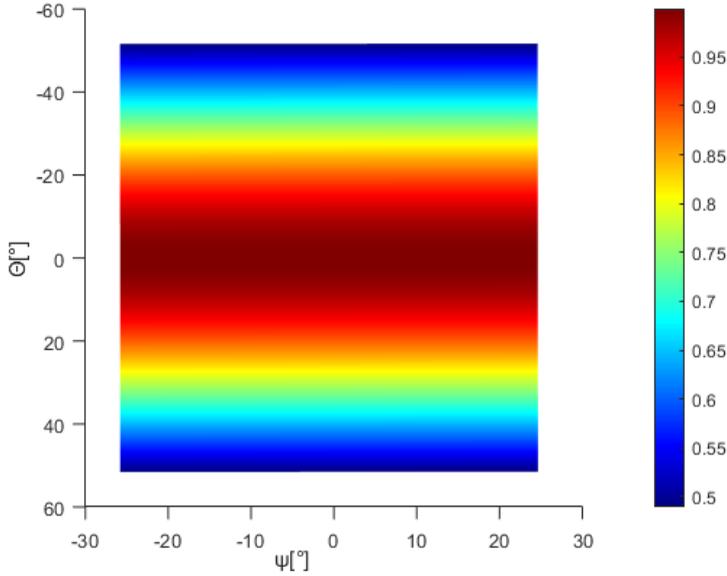

**Figure 7.** *Dexterity distribution of the serial chain in the* ($\psi$, $\theta$). *plane*

### 3.2. Parallel chain (PC)
#### 3.2.1. Kinematic model
The parallel chain of the nHH device is defined by a 3-RRR parallel planar manipulator. This latter is composed of three identical kinematic chains connecting a mobile platform to a fixed base. Each chain is consisting of an actuated revolute joint attached to the ground followed by two revolute joints to be connected to the platform, as show on Fig. 8.

The center of the joint connecting the two links of the $i^{th}$ chain will be referred to as $B_i$. The length of the links of the $i^{th}$ chain will be noted $L_1$ (for link $A_iB_i$) and $L_2$ (for link $B_iC_i$). The active and passive revolute joints are denoted by $\theta_i$ and $\beta_i$, respectively, with $i = 1,2,3$.

The position of the moving platform is defined by the coordinates of point $P(Px, Py)$ in the fixed reference frame $R_0$ and its orientation is given by the angle φ between one axis of the fixed reference frame and the corresponding frame R1 of the moving frame, as show on Fig.8.

The inverse geometric model, to obtain the active joints from the position and the orientation of the moving platform, can be expressed as follow[28]:

$$\theta_i = 2\ tan^{-1} \frac{-\bar{N}_i \pm \sqrt{\bar{M}_i^{\,2} + \bar{N}_i^{\,2} - \bar{L}_i^{\,2}}}{\bar{L}_i - \bar{M}_i} \tag{14}$$

where,

$\bar{M}_i = 2a_{ix}L_1 + 2L_1L_3\ cos(\varphi+\alpha_i)\ - 2P_xL_1$  (15)

$\bar{N}_i = 2a_{iy}L_1 + 2L_1L_3\ sin(\varphi+\alpha_i)\ - 2P_Y L_1$  (16)

$\bar{L}_i = a_{ix}^{\,2} + a_{iy}^{\,2} + P_x^{\,2} + P_y^{\,2} + L_1^{\,2} + L_3^{\,2} + 2a_{ix}L_3\ cos(\varphi+\alpha_i) - 2\ a_{ix}\ P_x - 2P_x\ L_3\ cos\ (\varphi+\alpha_i) + 2a_{iy}L_3\ sin(\varphi+\alpha_i) -\ 2\ a_{iy}\ P_x - 2P_y\ L_3\ sin(\varphi+\alpha_i) - L_2^{\,2}$.  (17)

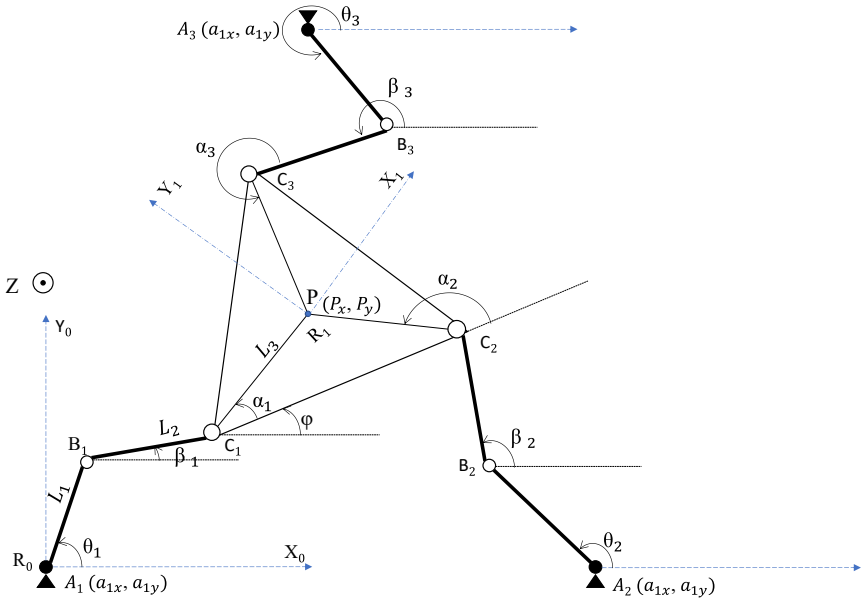

*Figure 8.* Geometric parameters of the parallel chain (3-RRR) of the nHH device.

The kinematic model of the 3-RRR parallel planar manipulator can be computed by the derivation of the forward kinematic model presented by the following equations [28]:

$$\mathbf{OP} = \mathbf{OA} + \mathbf{AB} + \mathbf{BC} + \mathbf{CP} \tag{18}$$

$$P_x = a_{ix} + L_1 \cos\theta_i + L_2 \cos\beta_i + L_3 \cos(\varphi + \alpha_i) \tag{19}$$

$$P_y = a_{iy} + L_1 \sin\theta_i + L_2 \sin\beta_i + L_3 \sin(\varphi + \alpha_i) \tag{20}$$

To finally obtain an equation expressed as follows:

$$\mathbf{J_\theta}\ \dot{\theta} =\ \mathbf{J_x}\dot{X}$$

where $\mathbf{J}_x$ and $\mathbf{J}_\theta$ are the parallel part and the serial part of the Jacobian matrix presented by Eq. (21) end (22), respectively.

$$\mathbf{J_x} = \begin{bmatrix} Jx_1 & Jy_1 & Jz_1 \\ Jx_2 & Jy_2 & Jz_2 \\ Jx_3 & Jy_3 & Jz_3 \end{bmatrix}; \qquad (21)$$

$$\mathbf{J_\theta} = \begin{bmatrix} J\theta_1 & 0 & 0 \\ 0 & J\theta_2 & 0 \\ 0 & 0 & J\theta_3 \end{bmatrix}; \qquad (22)$$

where,

$$Jx_i = 2P_x - 2a_{ix} - 2L_3 \cos(\varphi+\alpha_i) - 2L_1 \cos(\theta_i). \qquad (23)$$

$$Jy_i = 2P_y - 2a_{iy} - 2L_3 \sin(\varphi+\alpha_i) - 2L_1 \sin(\theta_i). \qquad (24)$$

$$Jz_i = 2L_3 a_{iy} \cos(\varphi+\alpha_i) - 2L_3 a_{ix} \sin(\varphi+\alpha_i) - 2L_3 P_y \cos(\varphi+\alpha_i) \qquad (25)$$
$$+ 2L_3 P_x \sin(\varphi+\alpha_i) - 2L_1 L_3 \sin(\varphi+\alpha_i)\cos(\theta_i) - 2L_1 L_3 \sin(\varphi+\alpha_i)\sin(\theta_i)$$

$$J\theta_i = 2L_1 P_x \sin(\theta_i) + 2L_1 a_{1y}\cos(\theta_i) - 2L_1 a_{1x}\sin(\theta_i) - 2L_1 P_y \cos(\theta_i) \qquad (26)$$
$$+ 2L_1 L_3 \sin(\varphi+\alpha_i)\cos(\theta_i) - 2L_1 L_3 \cos(\varphi+\alpha_i)\sin(\theta_i).$$

The Jacobian matrix satisfies Eq.27:

$$\mathbf{J_p} = \mathbf{J_\theta}^{-1}\mathbf{J}_x \qquad (27)$$

### 3.2.2. Dexterity

In order to evaluate the kinematic performance of the parallel chain of the nHH device, the dexterity will be examined. This criterion allows to measure how far the moving platform is far from singularity inside the workspace and one of the most considered criteria in literature[29]. The dexterity is computed using the inverse of the condition number described by Eq. (28).

$$\mu_P = \frac{1}{K(\mathbf{J_P})} \qquad (28)$$

where, $K(\mathbf{J_p}) = ||\mathbf{J_P}|| \cdot ||\mathbf{J_P}^{-1}||$

In order to evaluate the global dexterity inside a desired workspace, the global conditioning index will be considered and defined as follows[30]:

$$\mu_P^G = \frac{\sum_{i=1}^{N} \mu_{Pi}}{N} \qquad (29)$$

$N$ : discretization parameter of the desired workspace.

In order to identify the coupling between the serial and the parallel chain of the nHH device, the relations between the cartesian coordinates of the center of the moving platform of the parallel chain and the orientation of the serial chain are determined. As shown on Fig. 6, these relations are defined by equation (30) and (31):

$$\begin{cases} P_x = -H \tan\theta \cos\psi \\ P_y = -H \tan\theta \sin\psi \end{cases} \quad \begin{array}{c}(30)\\(31)\end{array}$$

For each orientation $\psi$ and $\theta$ leading to a position of the moving platform, the active angles $(\theta_1, \theta_2, \theta_3)$ can be computed using the inverse kinematic model presented in section 3.2. As consequence, the dexterity distribution of the parallel chain can be obtained by a simple mapping in the $(\psi, \theta)$ plane and given in the figures 9,10 and 11 for different values of the self-rotation.

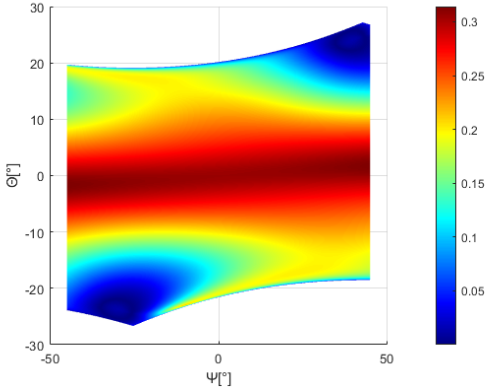

**Figure 9.** *Dexterity distribution of Parallel chain in the plane $(\psi, \theta)$ with self-rotation $\varphi = -20°$.*

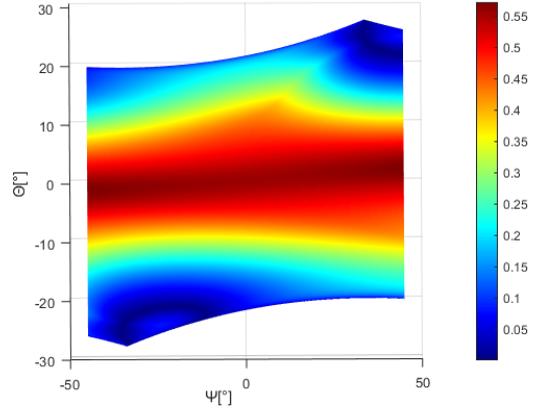

**Figure 10.** *Dexterity distribution of Parallel chain in the plane $(\psi, \theta)$ with self-rotation $\varphi = 0°$.*

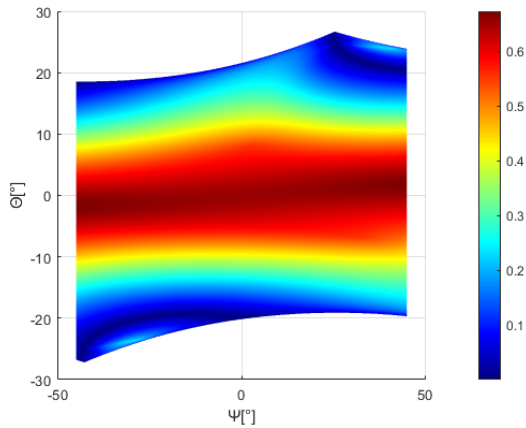

**Figure 11.** *Dexterity distribution of Parallel chain in the plane $(\psi, \theta)$ with self-rotation $\varphi = 20°$.*

## 4. Dimensional synthesis of the nHH device

The aim of this section is to identify the best design parameters of the nHH device leading to the best trajectory tracking accuracy. This latter can be characterized by the kinematic performance of the nHH device. The value of this performance depends on the positioning error, between the articulate coordinates and the operational ones, due to the kinematic transformations.

The optimization process, based on the dimensional synthesis, allows identifying the nHH device that is more suitable for the criteria selected by the user and in our case the kinematic performance. In the literature, several methods and indices can be found[29], [31], [32] To compute the kinematic performance of a structure, we chose the global dexterity defined by Gosselin [29] as it characterizes the isotropy of the robot velocity.

### 4.1. Problem formulation

This section focuses on the development and the results of the multidimensional, nonlinear optimization problem of selecting the geometric design variables for the nHH device having a specified workspace as well as the best dexterity distribution. The aimed application is the minimally invasive surgery due to the experimental recorded path used in the comparison study in section 5 but can be enlarged to others.

Fig. 12 presents the intersection between the cone and the plane crossing the parallel chain. This intersection is bounded by a circle $C$ and discretized in $n$ point $E_i$.

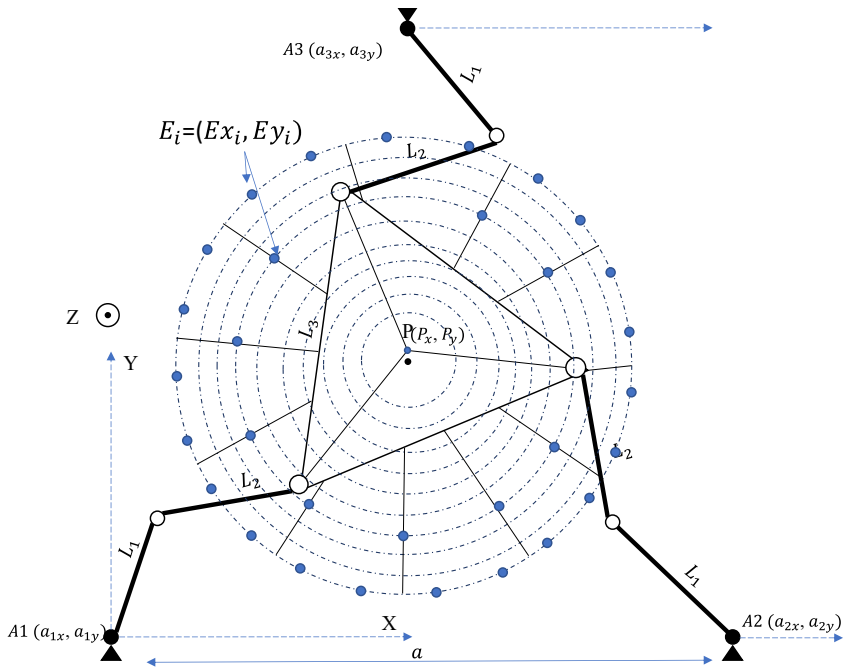

***Figure 12.*** *Intersection between the cone and the parallel chain plane.*

The proposed approach is based on the minimization of the objective function M(L) which reflects the overall performance of the manipulator by associating two separate indexes for the parallel chain and the serial chain with giving more weight to the parallel part, since the serial chain has a good performance even at the edges of the workspace. Indeed, the performance of the whole device is related to each chain and the performance of the device can reach a good level only when the performance of both parts is good. By this association, we manage to have a performance indicator of the whole manipulator. This approach is based on using a Genetic Algorithm (GA) method [33]–[35] .The optimization problem is stated as follows:

$$\text{Minimise } M(\mathbf{L}) = \sigma \cdot \bar{\mu}_P(\mathbf{L}) + (1-\sigma) \cdot \bar{\mu}_S(\mathbf{L}) \tag{32}$$

Subject to,

$$W_E = \sum_{j=1}^{3} \sum_{i=1}^{n} P_j(E_i) \leq 0 \tag{33}$$

With

$$P_j(E_i) = (E_{xi}\text{-}a_{ix})^2 + (E_{yi}\text{-}a_{iy})^2 - (L_1 + L_2 + L_2 - \delta)^2 \tag{34}$$

$$u_{k=\{1,5\}} = \{L_1 \quad L_2 \quad L_3 \quad a \quad H\}; \quad u_k \in [u_{k\,min};\, u_{k\,max}] \quad \mathbf{V} = [L_1 \quad L_2 \quad L_3 \quad a \quad H]^T$$

Where,
$\bar{\mu}_P(\mathbf{L}) = \frac{1}{N}\sum_{i=1}^{N}\mu_i(\mathbf{L})$ : global dexterity of the parallel manipulator.
$\bar{\mu}_S(\mathbf{L}) = \frac{1}{N}\sum_{i=1}^{N}\mu_i(\mathbf{L})$ : global dexterity of the serial chain.
$\sigma$ : weighting coefficient ($0 < \sigma < 1$). $\sigma = 0.7$.
$P_j(E_i)$: constraint of the existence of the workspace.
$M(\mathbf{L})$: Objective functions defining the criteria, the kinematic performance, to optimize.

**V**: Design vector of the parallel chain with the length $L_i$ ; $i \in \{1,2,3\}$, the distance $a$ representing the dimension of the base and the distance H between the PC plane and the CoR .

$\delta$ : the safety margin for singularity avoidance.

### 4.2. Genetic algorithm method, optimization and results
Genetic Algorithms (GAs) are heuristic search algorithms based on the mechanism of natural selection and natural genetics initially proposed by Holland [36]. They have been used in a variety of engineering fields such as in machine design. A real-coded GA[37]is used here to solve the optimization problem.

In the present application, each individual is a design vector, $\mathbf{L} = [L_1 \quad L_2 \quad L_3 \quad a \quad H]^T$. It corresponds to the nHH manipulator, and its characteristics are design parameters. A population of 50 individuals is manipulated through 300 generations.

The algorithm is allowed to select the optimal values of the design parameters in the bounding intervals given in the Table1.

*Table 1. Bounding intervals of the parallel chain design parameters.*

|         | $L_1$ | $L_2$ | $L_3$ | $a$ | $H$ |
|---------|-------|-------|-------|-----|-----|
| UB [mm] | 10    | 10    | 20    | 100 | 100 |
| LB [mm] | 100   | 100   | 70    | 300 | 250 |

The optimal design vector $\mathbf{L}_{op}$ obtained using the GA is presented by the following:

$$\mathbf{L}_{op} = [80 \quad 80 \quad 70 \quad 215 \quad 125]^T$$

The figure below represents the dexterity distribution of the parallel chain within its workspace in the plane $(X, Y)$ with a self-rotation of the moving platform equal to $0°$.

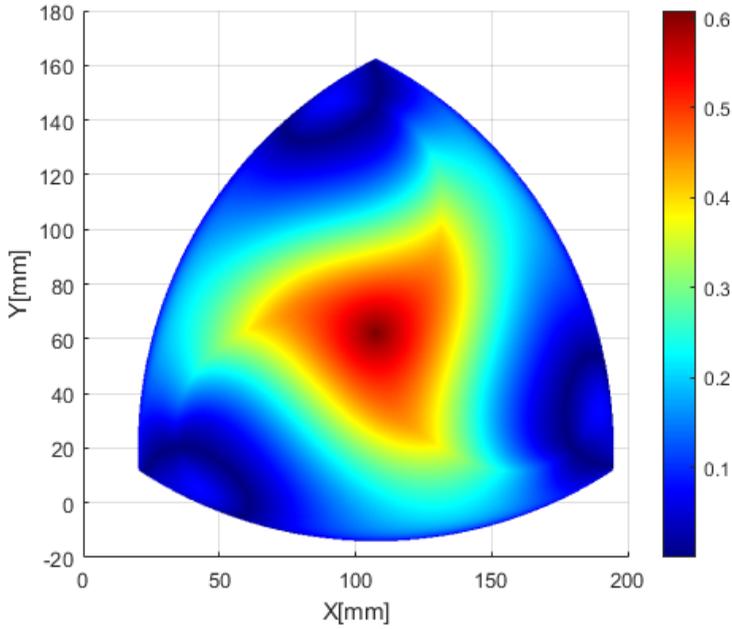

*Figure 13. Dexterity distribution of the PC at $\varphi = 0°$.*

The optimization of the serial chain is obtained by optimizing the distance H between plan of the parallel chain and the CoR. In fact, as mentioned in the section 2, the projection of the cone in the parallel chain plane yields to a circular border defined by $E_i$ (Fig. 13) with the radius linked to the distance H. The coordinates of the points $E_i$ are given by Eq. (35).

$$E_i = \begin{cases} E_{xi} = r.\cos\rho \\ E_{yi} = r.\sin\rho \end{cases} \text{ for } \rho = [0, : \pi/12 : 2\pi] \qquad (35)$$

With

$$r = 61mm$$

Figures 14 and 15 give a visual representation of the task workspace in the PC plane and 3D representation of the workspace with nHH device, respectively.

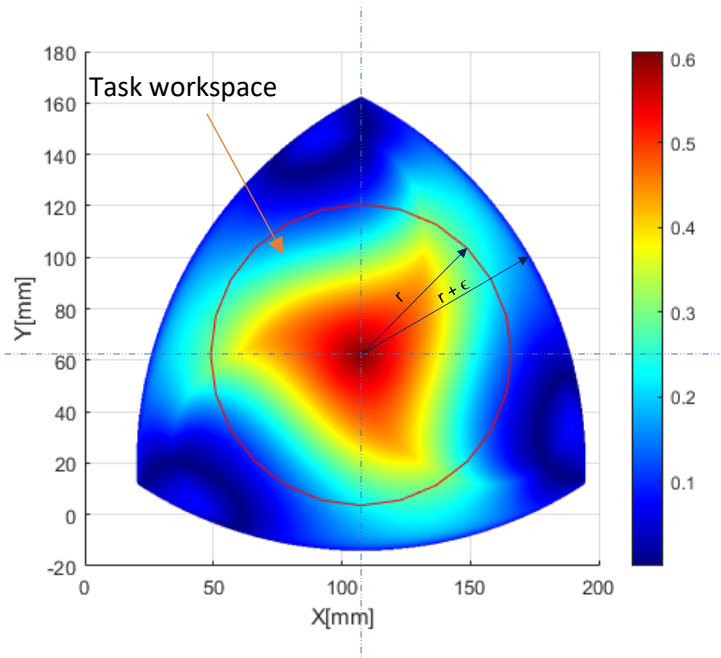

*Figure 14.* Task workspace in the PC plane.

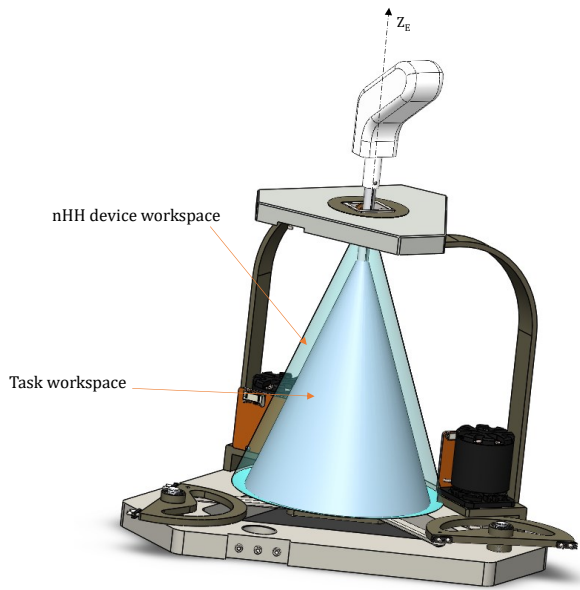

*Figure 15.* 3D representation of the task workspace and the nHH workspace.

## 5. Numerical validation of haptic feedback and comparison

The nHH device represents the master station of a master-slave platform. This device is used to control the slave robot and ensure the force feedback in case of the slave robot interaction with its external environment. The nHH device is designed using a mechanical solution based on a capstan associated with a simplex DC motor. A set of four capstans and four simplex DC motors can be observed on the prototype. The capstan for the revolute joint is with a ratio of 4.1 and the one for the translation is with a ratio of 3.4. Regarding the actuation, the DC motors present a nominal torque of 0.8 [Nm]. The first numerical validation is presented in this section and in future works, the experimental study will be developed and validated. So, in order to validate the kinematic performance of the optimal solution, in this section the model of the nHH device is developed using a SimMechanics model based on rigid-body dynamics according to the scenario presented in Figures 16 and 17. For different configuration $N(Nx, Ny)$ of the nHH as presented in Fig.18 we determine the feedback force on the end-effector by applying the torques on the actuated joints . In addition, a comparison study is performed to prove the efficiency of the proposed architecture of nHH device.

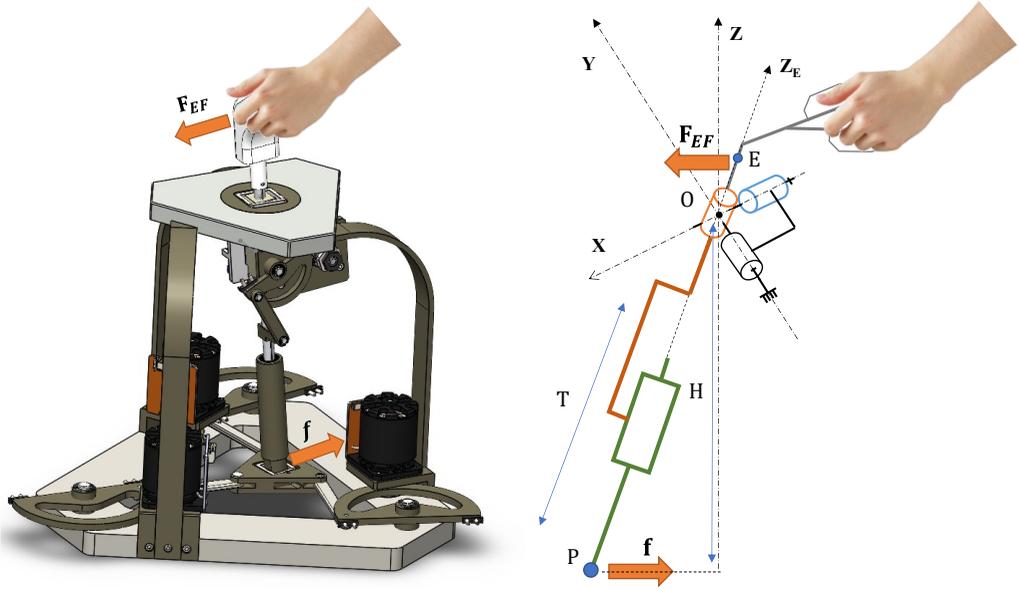

**Figure 16.** *Applied efforts on the nHH device.*

The nHH device is assumed to be a rigid body; therefore, deformations are not considered in this study. A function that calculates the actuated joint torques is constructed. The inputs to this function are the references forces $\mathbf{F}_{EF}$ and $\mathbf{f}$, which are the force vector applied on the end effector of the mobile parallel platform as presented in Fig.16, respectively.

The analytic method to compute the actuated joint torques of the parallel chain for a given reference force $\mathbf{f}$ is defined by the following relation :

$$\boldsymbol{\tau} = \mathbf{J}_P^T . \mathbf{f} \tag{36}$$

with

$\mathbf{f}$: force applied on the parallel mobile platform, $\mathbf{f} = [f_x \quad f_y \quad m_z]^T$.

$\mathbf{J}_P^T$ : Transpose of the Jacobian matrix of the parallel manipulator.

$\boldsymbol{\tau}$ : Actuated joints torque, $\boldsymbol{\tau} = [\tau_1 \quad \tau_2 \quad \tau_3]^T$ .

The input are the actuated joints torques calculated using the equation above. The output is the applied force on the end effector.

The relation between the two efforts is denoted by Eq.37 .The joint torque is inserted into the SimMechanics model to compute the force $\mathbf{F}_{EF}$, for different configurations.

$$\mathbf{OE} \wedge \mathbf{F_{EF}} = \mathbf{OP} \wedge \mathbf{f} \tag{37}$$

with

$$\mathbf{F_{EF}} = \begin{pmatrix} F_x \\ F_y \\ M_z \end{pmatrix}, \mathbf{f} = \begin{pmatrix} f_x \\ f_y \\ m_z \end{pmatrix}, \mathbf{OE} = \begin{pmatrix} x_E \\ y_E \\ z_E \end{pmatrix}, \mathbf{OP} = \begin{pmatrix} x_p \\ y_p \\ z_p \end{pmatrix};$$

For $M_z = m_z = 0$, the relation between the applied effort on the end effector and the mobile platform is given by the equations below :

$$\begin{cases} F_x = \dfrac{z_p}{z_E} \cdot f_x \\ F_y = \dfrac{z_p}{z_E} \cdot f_y \end{cases} \tag{38}$$

$$\tag{39}$$

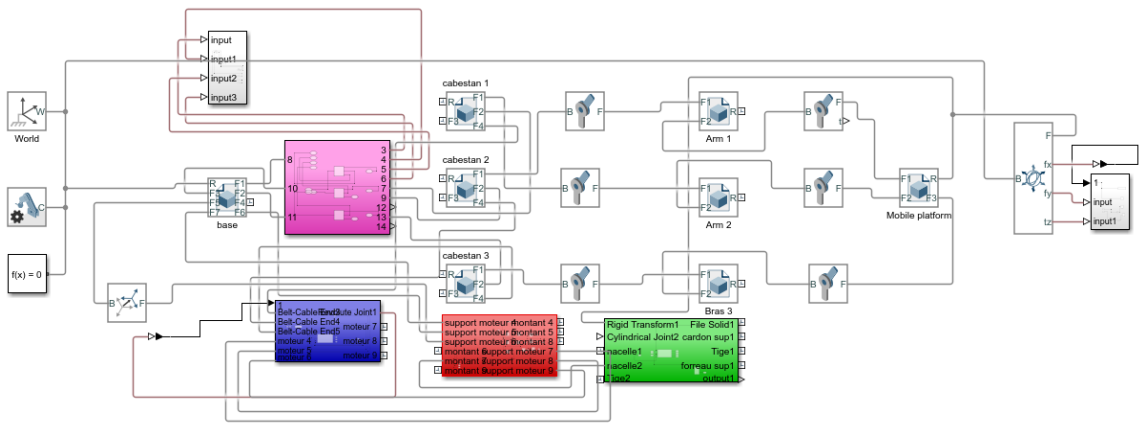

*Figure 17. Validation model: Simscape model of the nHH device.*

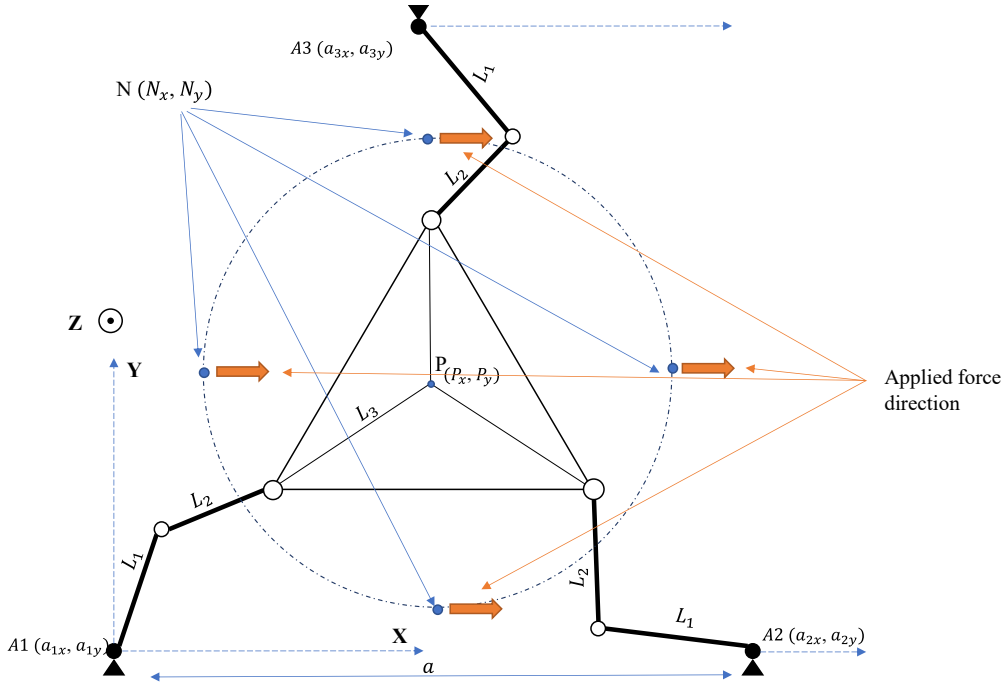

*Figure 18. Configurations of nHH.*

Table 2 resume the applied torques, analytic force value and the simulation force value on the end effector for the four different configurations $N(Nx, Ny)$ at the edges of the workspace as presented in Fig.18 $N(Nx, Ny)$. Both the analytic and the simulation forces where calculated using the serial chain and the parallel chain of the nHH device.

*Table 2. Force values depending on nHH configuration.*

|  | $Nx$[M] | $Ny$[M] | ANALYTIC FORCE VALUE[N] | SIMULATION FORCE VALUE[N] |
|---|---|---|---|---|
| **POSITION 1** | 0.121 | 0.030 | 1 | 0.98 |
| **POSITION 2** | 0.121 | 0.083 | 1 | 1.0065 |
| **POSITION 3** | 0.140 | 0.075 | 1 | 1.021 |
| **POSITION 4** | 0.095 | 0.075 | 1 | 0.94 |

As follows from the table shown above, we manage to prove that the proposed HH device has a good force feedback and capable to regenerate the force applied by the surgeon in order to stimulate the real and the virtual environment. Despite the fact that the chosen points where we calculate the force feedback are at the edges of the workspace (with an altered dexterity index), we still manage to have suitable feedback where the error does not exceed 2% for a force applied along the X axis as presented in Figure 18.

The next a comparison between the nHH and the SPM[12], [38] devices is performed. The spherical parallel manipulator, as the nHH device, is meant to be used as a haptic device in robotic system for minimally invasive surgery application. In order to prove the efficiency of the proposed architecture of nHH device, a comparison study is proposed[39].

The comparison of the two manipulators is based on the capability of each device to perform motions around a specific point in its workspace. Dexterity, which represent the amplification of errors as a result of kinematic and static transformation between cartesian and joints spaces, is chosen to evaluate the kinematic for each manipulator. For the SPM, the dexterity has been calculated and determined by Saafi [13], [23].

Based on results in [23], the maximum value of the dexterity that can be reached is about 0.4 at the center of the workspace with a self-rotation $\varphi = 0°$.

The comparison is based on choosing a recorded trajectory in the workspace of both manipulators, then calculate the dexterity along this path. The trajectory has been obtained during a recorded surgeon motion in a real mini-invasive surgery task using a motion capture system[24]. Figure 19 gives a graphical representation of the trajectory in the plane $(\psi, \theta)$.

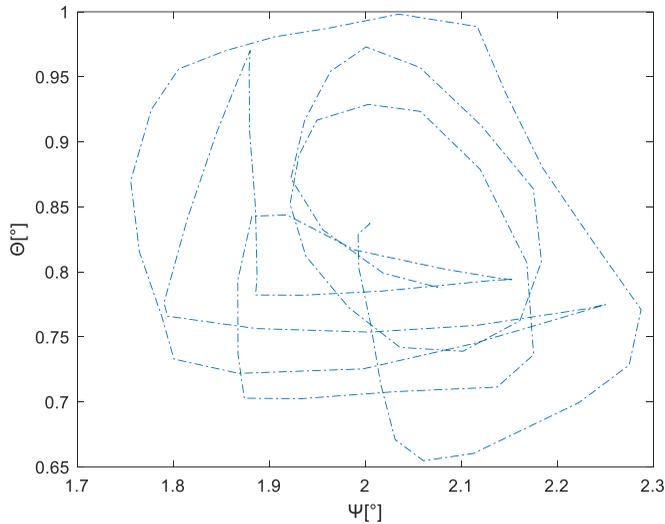

***Figure 19.*** *Trajectory obtained from a recorded surgeon gesture by motion capture.*

The same trajectory was implemented on the nHH device. The obtained results for different values of self-rotation, $\varphi = 0°$, $\varphi = 50°$ and $\varphi = -10°$, are shown on Figures 20, 21 and 22, respectively.

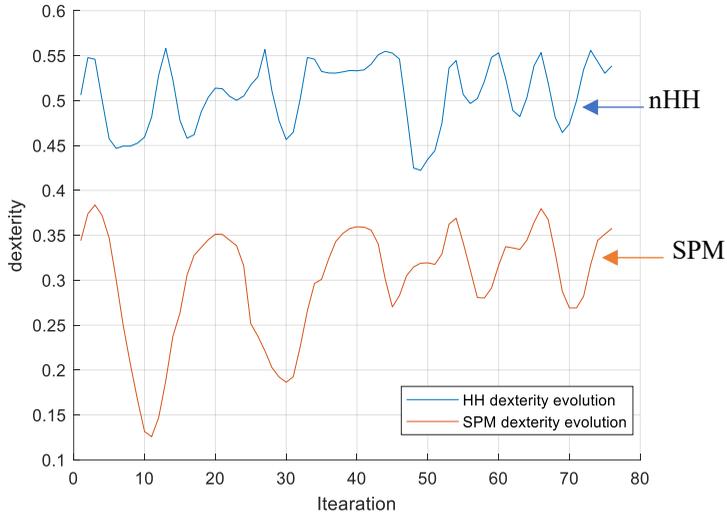

***Figure 20**. Dexterity evolution for nHH and SPM for φ=0°.*

Table.3 resumes the maximum, minimum and mean dexterity along the chosen trajectory for both manipulators with a self-rotation *φ=0°*.

***Table 3.** Dexterity values for SPM and nHH device for φ=0°.*

|  | MAX | MIN | MEAN |
|---|---|---|---|
| **SPM** | 0.3838 | 0.1258 | 0.3001 |
| **nHH** | 0.5581 | 0.4223 | 0.5071 |

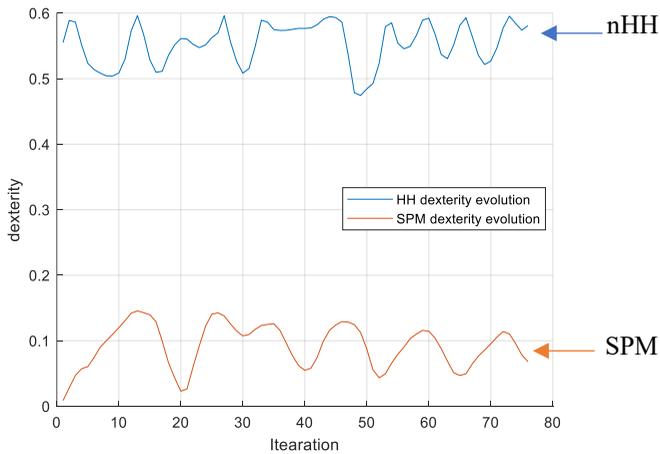

***Figure 21**. Dexterity evolution for nHH and SPM for φ=50°.*

Table.4 resume dexterity values along the chosen trajectory for both manipulators with a self-rotation $\varphi=50°$.

**Table 4.** dexterity values for SPM and nHH device for $\varphi=50°$.

|  | MAX | MIN | MEAN |
|---|---|---|---|
| **SPM** | 0.1455 | 0.0087 | 0.0924 |
| **HH** | 0.5962 | 0.4744 | 0.5536 |

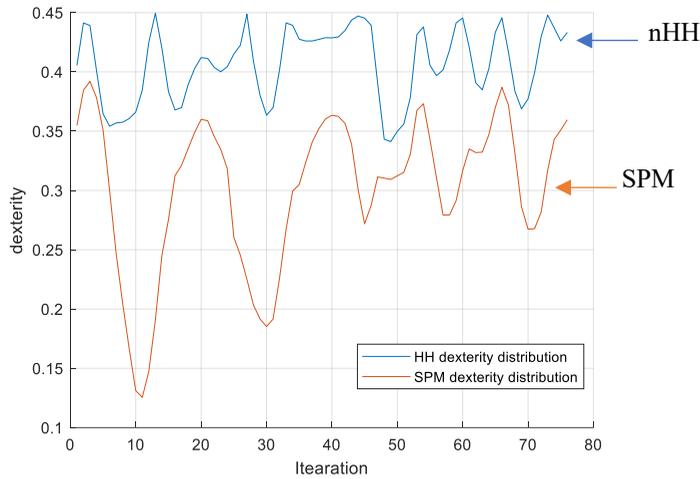

**Figure 22.** Dexterity evolution for HH and SPM for $\varphi=-10°$.

Table.5 resume dexterity values along the chosen trajectory for both manipulators with a self-rotation $\varphi=-10°$.

**Table 5.** Dexterity values for SPM and HH device for $\varphi=-10°$.

|  | MAX | MIN | MEAN |
|---|---|---|---|
| **SPM** | 0.3920 | 0.1255 | 0.3016 |
| **HH** | 0.4495 | 0.3412 | 0.4066 |

For a self-rotation $\varphi=0°$ and along the chosen trajectory the SPM's dexterity goes from 0.38 to 0.12. This is explained by the presence of singularities inside the workspace which alters the kinematic performance and amplifies the errors. Moreover, near singular configurations the self-rotation is no more controllable. However, the nHH device represents a good dexterity along the same trajectory which can reach 0.55 at it's maximum.

Consider Fig.21, which plots the SPM's dexterity against the nHH's dexterity for a self-rotation $\varphi=50°$. From the resulting plot, we can confirm that the SPM have a bad kinematic performance. The new haptic device presents a maximum dexterity of 0.59. This value is greater than the SPM's which is limited to 0.3.

According to the results presented above by the three figures and the three tables, the suitability and potential of the novel architecture as a haptic interface for surgical tasks. Since the dexterity of the nHH device is above 0.3412 for the different orientations and along the chosen trajectory, the new hybrid manipulator is adequate for many tasks, especially for minimally invasive surgical ones.

## 6. Conclusion

In this paper we proposed a new hybrid haptic (nHH) device with a fixed center of rotation and 4-doF. The nHH is designed to be used as a master device in robotic platform dedicated to laparoscopic surgery. The proposed device is an association of two chains, a serial and a parallel chain. The serial chain (SC) allows to handle the translational motion (1-dof) and the parallel chain (PC) allows to handle the rotational motion (3-dofs), mainly the two tilt motions and the self-rotation.

The geometric parameters of the nHH device were optimized in order to fit the task workspace as well as a best kinematic performance distribution over its workspace. Global dexterity has been chosen as a criterion to characterize this kinematic performance. A safety margin distance is included in the optimization to eliminate singular region from the workspace. A kinematic model of the nHH device has been developed and validated using a SimMechanics model.

A comparison study has been performed between the proposed device and a spherical parallel manipulator (SPM) dedicated to a similar application. A real surgeon gesture is considered for the orientation of the master device. The results obtained prove that the nHH device presents a more interesting behavior then the SPM and still far from singularity. The global dexterity index is over 0.5 and its value is not altered by the self-rotation which is one of the major limitations of the SPM.

The CAD model and first prototype of the nHH device are presented and will be used in future work which will be focused on the master-slave scheme and haptic control implementation.

**Authors' Contributions.** M.M., H.S., A.M., M.A. and M.A.L. conceived and designed the study. M.M., M.A. and M.A.L. conducted and analyzed the numerical experiments. M.M., H.S., A.M., M.A., S.Z. and M.A.L. wrote the article.

**Financial support.** This work was financially supported by the "IRP-RACeS" program IRP : Robotic Assisted System for Safe Cervical Surgery supported by the CNRS in France and European program of international students' mobility.

**Ethical considerations.** None.

**Conflicts of interest.** The authors declare no conflict of interest.